\documentclass[journal=jacsat,manuscript=article]{achemso}

\usepackage[version=3]{mhchem} 
\usepackage[T1]{fontenc}
\usepackage{xcolor}

\author{Arshad Mehmood}
\affiliation[DoIT]{Division of Information Technology - Research Computing, Informatics \& Innovation, Stony Brook, New York 11794, USA} 
\altaffiliation{Institute for Advanced Computational Science, Stony Brook University, Stony Brook, New York 11794, USA}

\author{Caitlin V. Hetherington}
\altaffiliation{Institute for Advanced Computational Science, Stony Brook University, Stony Brook, New York 11794, USA}
\affiliation[Chemistry]
{Department of Chemistry, Stony Brook University, Stony Brook, New York 11794, USA}

\author{Zain Zaidi}
\altaffiliation{Institute for Advanced Computational Science, Stony Brook University, Stony Brook, New York 11794, USA}
\affiliation[Chemistry]
{Department of Chemistry, Stony Brook University, Stony Brook, New York 11794, USA}

\author{Benjamin G. Levine}
\email{ben.levine@stonybrook.edu}
\altaffiliation{Institute for Advanced Computational Science, Stony Brook University, Stony Brook, New York 11794, USA}
\affiliation[Chemistry]
{Department of Chemistry, Stony Brook University, Stony Brook, New York 11794, USA}

\title[An \textsf{achemso} demo]
  {Chemical and Conformational Control of the Spectroscopic Properties of Multi-Layer and Multi-Defect Carbon Dots}

\abbreviations{IR,NMR,UV}

\begin{document}

\begin{tocentry}
\includegraphics[width=8.25cm]{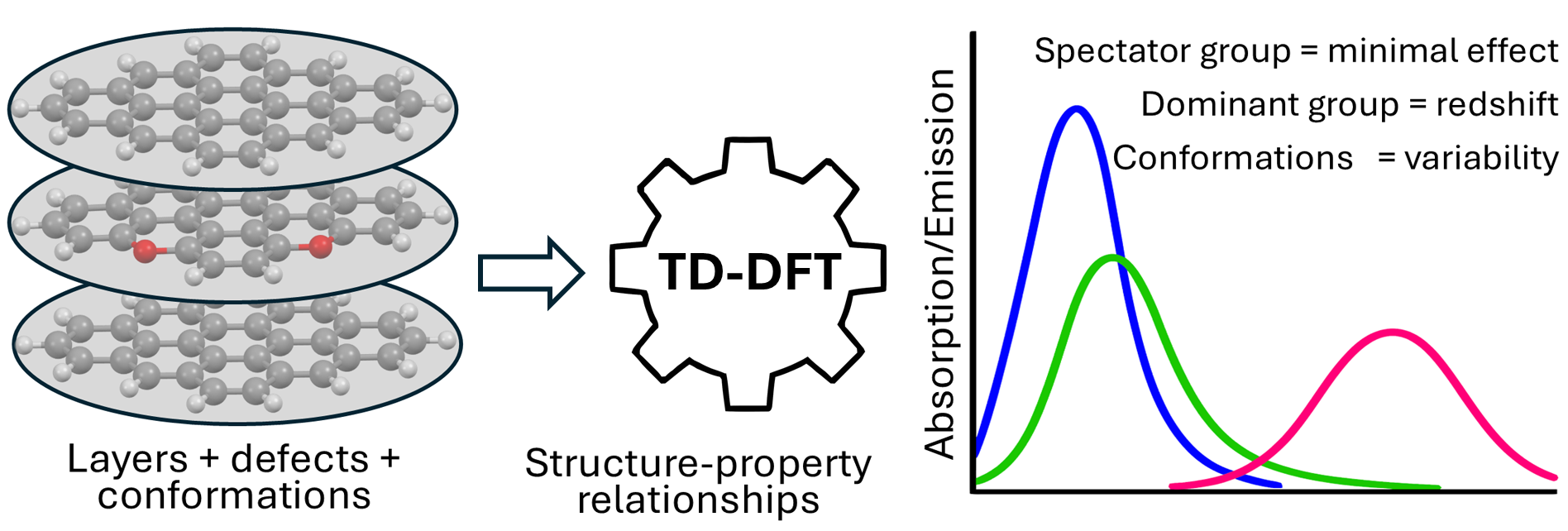}
\end{tocentry}

\begin{abstract}
Carbon dots (CDs) are renowned for their bright and tunable photoluminescence (PL), stability, and biocompatibility, yet it remains challenging to link their heterogeneous structures to their spectroscopic properties. This study utilizes density functional theory (DFT) and time-dependent DFT (TD-DFT) to systematically investigate how the spectroscopic properties of complex CDs with multiple layers and multiple defects are determined by their structures and compositions. Calculations reveal that strongly oxidizing defects, such as carbonyl and carbonyl acetate, significantly redshift absorption and emission spectra. In contrast, less oxidizing defects, such as hydroxyl, behave as spectators with minimal impact on absorption and emission, except when they interact strongly with more oxidizing defects. We find that not only the excitation energy but also the excitation character itself is impacted by the presence of specific defects, and the pH-dependence of the spectroscopic properties can be attributed to their protonation state-dependent excitation character. We show that the twisting, sliding, and linker-mediated folding of surface-functionalized layers in CDs markedly alter excitation energies and characters, offering a molecular explanation for experimentally observed emission intermittency and polarization fluctuations. These insights provide strategies for optimizing CDs for various applications, including bioimaging, photocatalysis, and optoelectronic devices.

\end{abstract}

\section{Introduction}
Carbon dots (CDs) represent an emerging class of zero-dimensional carbon-based nanomaterials, characterized by their small particle size (less than 10 nm) and unique structural and functional properties.\cite{Zhao2024, Zhou2022, Yu2021, Carbonaro2024, Madonia2023} Their low cost synthesis, high conductivity, remarkable stability, and ease of surface modification have attracted considerable scientific interest. \cite{Yu2021, Lim2015, Li2012} One of the most compelling features of CDs is their tunable photoluminescence (PL), which, combined with their high biocompatibility and low toxicity, positions them as ideal candidates for a variety of applications, including bioimaging, drug delivery, photocatalysis, and optoelectronic devices.\cite{Zhao2024, Zhou2022, Yu2021, Lim2015, Li2012, Hola2014} 
\par
Generally, the core–shell-like structure of CDs consists of a base carbon core with chemical functional groups attached the surface.\cite{Yuan2018, Langer2021, Carbonaro2019, Yu2021, Lim2015, Li2012}Carbon cores consist of sp$^2$-hybridized graphene fragments separated by sp$^3$-hybridized defects and can incorporate heteroatoms such as nitrogen, phosphorus, or oxygen.\cite{Tepliakov2019, Bian2023, Sk2014, Yoon2016, Kundelev2022, Feng2019, Olla2023} These heteroatoms significantly influence the size of the sp$^2$-hybridized subdomains within the graphitic carbon core, leading to photoluminescence behavior that is less dependent on particle size than expected from simple quantum-confinement arguments. \cite{Bian2023, Mandal2023, Yoon2016, Zhang2023} Recent experimental studies have further shown that surface functionalization and heteroatom incorporation modulate absorption features ($\pi$–$\pi$* vs. n–$\pi$*), the relative intensity and spectral weight of blue versus green emission bands, and excited-state lifetimes.\cite{Olla2023, Madonia2023, Szapoczka2024, Carbonaro2024} The surface of CDs typically includes common functional groups like amino, epoxy, carbonyl, aldehyde, hydroxyl, and carboxylic acid.\cite{Li2015, Yu2021, Zhao2024, Zhou2022} Additionally, CDs may contain other molecular fragments that contribute to their functionality, creating a high level of complexity in their structures.\cite{Yu2021, Bian2023}
\par
The PL behavior of CDs is intricately influenced by variations in their structure, size, doping, and environmental conditions, resulting in a complex and contradictory understanding of their emissive properties.\cite{Langer2021, Han2022, Liu2019, Zhao2024, Yu2021, Strauss2014, Nguyen2020, Lu2017} A major challenge lies in the limited control over these parameters, which contributes to ongoing debates regarding the origin of emission in CDs. Diverse synthetic strategies yield a broad spectrum of structural features, leading to significant variability in optical responses and inconsistent experimental observations.\cite{Jiang2015, Righetto2020, Qu2020, Essner2018, Duan2020, Song2015} This heterogeneity complicates efforts to develop a unified explanatory framework, with proposed emissive origins attributed to core structures, surface states, or molecular fragments. In general, the conjugated carbon core supports $\pi$–$\pi$* transitions,\cite{Sudolská2015, Yoon2016, Sk2014, Kundelev2022, Bai2023, Zhang2023} while surface-localized sp$^3$ hybridization can enable both $\pi$–$\pi$* and n–$\pi$* transitions at lower energies.\cite{Hu2013, Hola2014, Hola2017, Shao2019, Hazem2018, Zhao2024, Yu2021} Two principal strategies are employed to tune emission wavelengths: modifying the size of sp$^2$ domains within the core and engineering surface functionalities.\cite{Wang2021, Sun2006, Kundelev2019, Qu2013,Umami2022} Variation in core size determines the effective optical gap, while surface groups can further tune the energetics of the quantum-confined states and/or introduce new electronic states at the interface. Additionally, the presence of molecule-like fluorophores within some CDs introduces further complexity to their emissive behavior.\cite{Righetto2020, Yu2021, Bian2023, Qu2020} 
\par
An additional layer of complexity arises from environmental factors and structural flexibility. Protonation states, interlayer stacking, and conformational fluctuations can dramatically alter orbital alignment, dipole moments, and transition character.\cite{Siddique2020,Liu2016,Chen2020, Feng2017, Langer2023, Nguyen2020, John2024, Guo2024, Dilshener2024, Yadav2023, Minervini2025} As a result, the same CD may exhibit different photophysical signatures depending on its protonation environment or conformational state. This dynamic behavior is not only of fundamental interest but could also be harnessed for responsive or switchable optical materials.
\par
Electronic structure simulations have played a pivotal role in advancing the understanding of CDs, particularly their complex photoexcitation and PL properties and how these are modulated by structural and chemical factors. Numerous studies have examined the individual effects of surface functional groups, system size, shape, chemical doping, and environmental conditions on the optical spectra of CDs.\cite{Kundelev2022, Sk2014, Bai2023, Kundelev2019, Tepliakov2019, Kundelev2020, VuNhat2024, Chen2020, John2024, Kiani2024, Zhao2014, Zhao2024, Zhang2023, Yang2020, Niu2016,Jabed2021, Langer2023,Abdelsalam2018, Cao2021, Tuchin2024, Sudolská2015,Feng2017,Shtepliuk2018} Recent theoretical work has begun to elucidate the role that the flexible multi-layer/multi-defect structure of CDs plays in their optical properties.\cite{Sarkar2016,Umami2022,Santika2022,Siddique2020,Bian2023}  For instance, nitrogen and oxygen dopants are found to not only modify the electronic density but also influence conjugation length and symmetry, leading to excitation-dependent emission or inhomogeneous broadening.\cite{Bian2023, Righetto2020, Jabed2021, John2024} Furthermore, certain surface groups can serve as donor or acceptor centers, promoting charge transfer excitations that compete with intrinsic fluorescence pathways.\cite{Abdelsalam2018, Umami2024, Cao2021, Kundelev2020}
\par
In this work, we continue to build toward a unified understanding of structure-function relationships in these structurally messy materials, focusing on three related questions.  First, how does the combined presence of multiple surface groups and dopants influence the nature and energy of the lowest singlet excitations? Second, under what conditions do different excitation types such as localized $\pi$–$\pi$*, interlayer charge-transfer, or n–$\pi$* transitions dominate the excited-state manifold? Third, can realistic conformational changes such as interlayer rotations lead to abrupt changes in excitation character or energy?
To this end, we start by reestablishing a baseline understanding of the behaviors of individual defects at a consistent level of time-dependent density functional theory (TD-DFT).  From this baseline, we systematically study several series of model systems with different numbers of defects, protonation state, and particle size.  Finally, we investigate whether inter-layer dynamics can explain experimentally-observed fluorescence intermittency.  At all points, attention is paid to the relationship between intra- and inter-layer structure, excited state energetics, and excitation character.

\section{Computational Details}
Except where otherwise noted, the model CD system consists of three layers stacked on top of each other, with each layer comprising multiple fused benzene rings, either with or without dopant atoms. The central layer was decorated with various types of oxygen and nitrogen-containing surface defects. The model systems underwent ground-state geometry optimization using DFT with the CAM-B3LYP\cite{Takeshi2004} long-range corrected hybrid exchange-correlation density functional, incorporating the D3 empirical dispersion correction (Grimme D3)\cite{Grimme2010}, and the 6-31G(d,p) Pople basis set.\cite{Krishnan1980} The geometries and electronic structures were optimized without applying any constraints other than the electronic state's multiplicity. Given our interest in absorption and luminescence we focus exclusively on singlet states.  Excitation energies and oscillator strengths were calculated using the linear-response TD-DFT approach with the same DFT functional. We also evaluated several additional DFT functionals for calculating excitation energies. The results are presented in the supporting information Figure S1. 
\par 
To determine the radiative transitions and energies, the geometries of the model systems were optimized in the first singlet excited state (S$_1$). To assess solvent effects, additional absorption and emission calculations were performed using implicit solvent models (PCM) for toluene, methanol, acetonitrile, and water. These solvents span a wide range of dielectric constants to evaluate environmental screening effects. Throughout, we use excitation energy to denote the lowest vertical S$\textsubscript{0}$\textrightarrow S$\textsubscript{1}$ energy (the optical gap) and emission energy for S$\textsubscript{1}$\textrightarrow S$\textsubscript{0}$ at the S$\textsubscript{1}$ minimum; we avoid "band gap" for these finite molecular models except when explicitly referring to a HOMO–LUMO gap at the S$\textsubscript{0}$ geometry. All calculations use the TeraChem software package.\cite{Seritan2021, Seritan2020, Ufimtsev2009, Isborn2011}
\par
The choice of basis set is known to have a profound impact on the computed electronic and optical properties of CDs and related nanocarbon systems, particularly for excited states where diffuse functions can be important for describing charge-transfer and Rydberg-like transitions.\cite{Mocci2022} To evaluate this effect, benchmark TD-DFT calculations were performed for representative systems using basis sets containing diffuse functions (6-31+G(d,p), def2-SVPD, and aug-cc-pVDZ) and were compared with results obtained using 6-31G(d,p). The inclusion of diffuse functions resulted in systematic redshifts in the calculated absorption and emission energies; however, all qualitative trends remained unchanged. For most systems, the shifts were modest (mean absolute shift $\approx$ 0.06 eV for both absorption and emission), whereas a larger shift ($\approx$ 0.18 eV) was observed for the quinone containing defect with the def2-SVPD basis. Excluding quinone defect, aug-cc-pVDZ yielded the largest average absolute deviations among the diffuse bases tested ($\approx$ 0.07 eV for absorption and $\approx$ 0.06 eV for emission). These benchmark results (Figure S2) indicate that the 6-31G(d,p) basis set provides qualitatively reliable predictions and a consistent description of the optical behavior of the studied systems, while maintaining an efficient balance between computational cost and accuracy.

\section{Results and discussion}
\subsection{Influence of Surface Functionalization}
\begin{figure*}[ht]
 \centering
 \includegraphics[height=6.35cm]{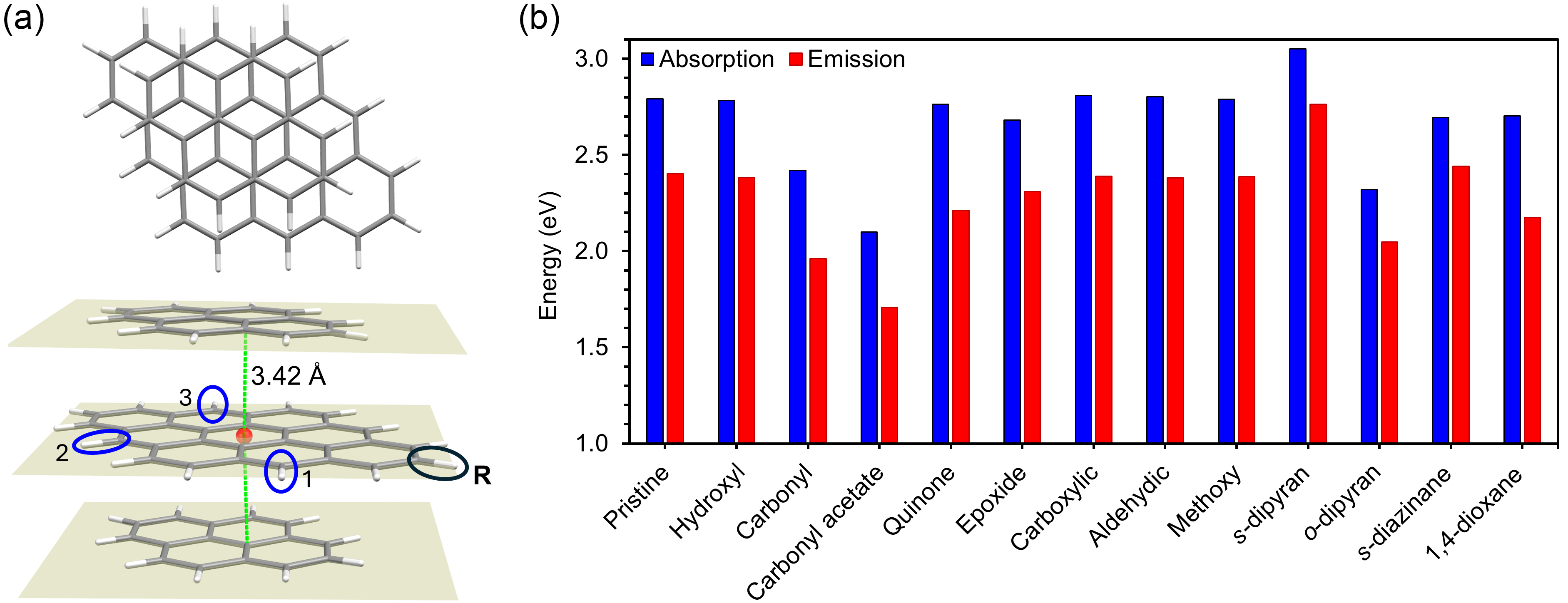}
 \caption{(a) The top and the side view of the model (idealized) three-layered carbon dot system used to study the effect of oxygen and nitrogen-containing surface defects. The outer layers are 3.42 Å apart from the central layer, as measured by the distance between planes passing through the atoms of each outer layer and the centroid position (red sphere) of the central layer. Surface defects are introduced by substituting the –CH groups (circled in black) of the fused aromatic ring in the central layer with common functional groups (R). The position marked using blue circles are substituted with oxygen atoms or –NH groups to obtain \textit{s}-/\textit{o}-dipyran or \textit{s}-diazinane defects (b) Calculated vertical excitation (S$\textsubscript{0}$\textrightarrow S$\textsubscript{1}$ transition) and emission energies of the model CD system decorated with different surface defects. The optimized coordinates of the systems are provided in the Supporting Information.}
 \label{fgr:Figure1}
\end{figure*}

CDs exhibit intricate architectures with diverse edge and potential in-plane functional groups, including oxygen, nitrogen, sulfur, and other heteroatoms. To systematically analyze the impact of various surface defects, we begin with a basic CD model incorporating a single oxygen- or nitrogen-containing modification. The considered defects include common functional groups such as hydroxyl, carboxyl, carbonyl, epoxide, formyl (aldehydic), and methoxy groups, as well as more complex heteroatomic defects in which oxygen or nitrogen atoms are embedded directly within the aromatic lattice, such as carbonyl acetate, quinone, 1,4-dioxane, and fused O- or N- heterocycles. The pyran defect is generated by replacing –CH groups with in-ring oxygen atoms, allowing the lattice to relax into a six-membered O-heterocycle (chromene, or benzopyran) fused to the $\pi$-system. When two neighboring pyranic units occur on the same side of the decorated layer (positions 1 and 2 in Figure \ref{fgr:Figure1}), the resulting defect is labeled \textit{s}-dipyran (same-side). This structure consists of two adjacent benzopyran rings fused to the aromatic edge. When the two pyran units lie on opposite sides (positions 1 and 3), the defect is referred to as \textit{o}-dipyran (opposite-side). Analogously, the \textit{s}-diazinane defect is obtained by replacing adjacent carbon atoms with sp$^3$-hybridized nitrogen atoms (–NH–), forming a six-membered diazinane ring fused to the core. The carbonyl-acetate defect combines a carbonyl and an acetate unit positioned on two neighboring aromatic rings. Other edge defects are introduced by substituting hydrogen atoms at the sites marked R in Figure \ref{fgr:Figure1} with the functional groups listed above. To avoid open-shell configurations that can arise from single-atom substitutions, adjacent carbon atoms were also modified to ensure a closed-shell singlet ground state. The resulting model comprises three stacked layers with the functionalized or defective layer positioned centrally (Figure \ref{fgr:Figure1}). In the optimized, defect-free configuration, the interlayer distance is 3.42 Å, which is comparable to the 3.34 Å spacing observed in graphite.\cite{Gulhan2019} Edge-bound oxygen defects, such as epoxides, exhibit behavior distinct from basal-plane variants,\cite{Chen2019, Feng2017, Sudolská2015} and we therefore restrict our analysis to edge functionalization. While the multilayer aromatic fragments used here represent idealized models, this framework has previously been shown to reproduce experimental emission line shapes with excellent fidelity\cite{Gomez2025}, indicating that it captures the essential photophysical behavior of CDs. Yet they certainly lack the structural disorder that is characteristic of experimentally realistic CD systems.  The goal of the present study is to develop structure-functional relationships from idealized models, but in future work, we hope to develop techniques for generating more realistic structural ensembles.  
\par
Analysis of absorbance and emission spectra for various surface defects (Figure 1b) reveals that the defects fall into two distinct groups. The \emph{spectator group} includes hydroxyl, aldehydic, quinone, epoxide, carboxylic, methoxy, \textit{s}-dipyran, \textit{s}-diazinane and 1,4-dioxane.  These defects produce spectra shifted only slightly compared to those of the pristine system, indicating minimal alteration of the underlying electronic structure. In contrast, the \emph{dominant group}, which includes carbonyl, carbonyl acetate, and \textit{o}-dipyran, exhibits pronounced spectral shifts. For example, carbonyl acetate causes a redshift of 0.69 eV in both the first absorption and emission bands.\cite{Reckmeier2016} Similarly, the carbonyl defect induces redshifts of 0.37 eV and 0.44 eV in the absorption and emission bands, respectively. These findings indicate that these dominant surface defects exert a significantly stronger influence on the optical properties of CDs than others. To further examine how the surrounding medium and dielectric screening affect these optical trends, we recalculated the absorption and emission energies using implicit solvent models (PCM) for toluene, methanol, acetonitrile, and water. The calculated spectra (Figure S3) show that solvent screening produces only minor changes in transition energies for most systems. The largest solvent-dependent shift is observed for the quinone defect, where both absorption and emission redshift by $\approx$ 0.30 eV from gas to water. This redshift is accompanied by a change in excitation character from a HOMO→LUMO+1 transition in the gas phase to a HOMO→LUMO transition in polar solvents. In contrast, the carbonyl acetate defect shows a blue-shift of $\approx$ 0.13 eV (absorption) and $\approx$ 0.18 eV (emission) from gas to water, indicating a slight stabilization of higher-energy transitions in polar solvents. For all other oxygen- and nitrogen-containing defects, solvent-induced variations stay within 0.07 eV, indicating that the intrinsic defect-dependent optical trends discussed above are largely preserved upon solvation. 
\par
\begin{figure*}[ht]
 \centering
 \includegraphics[height=12.0cm]{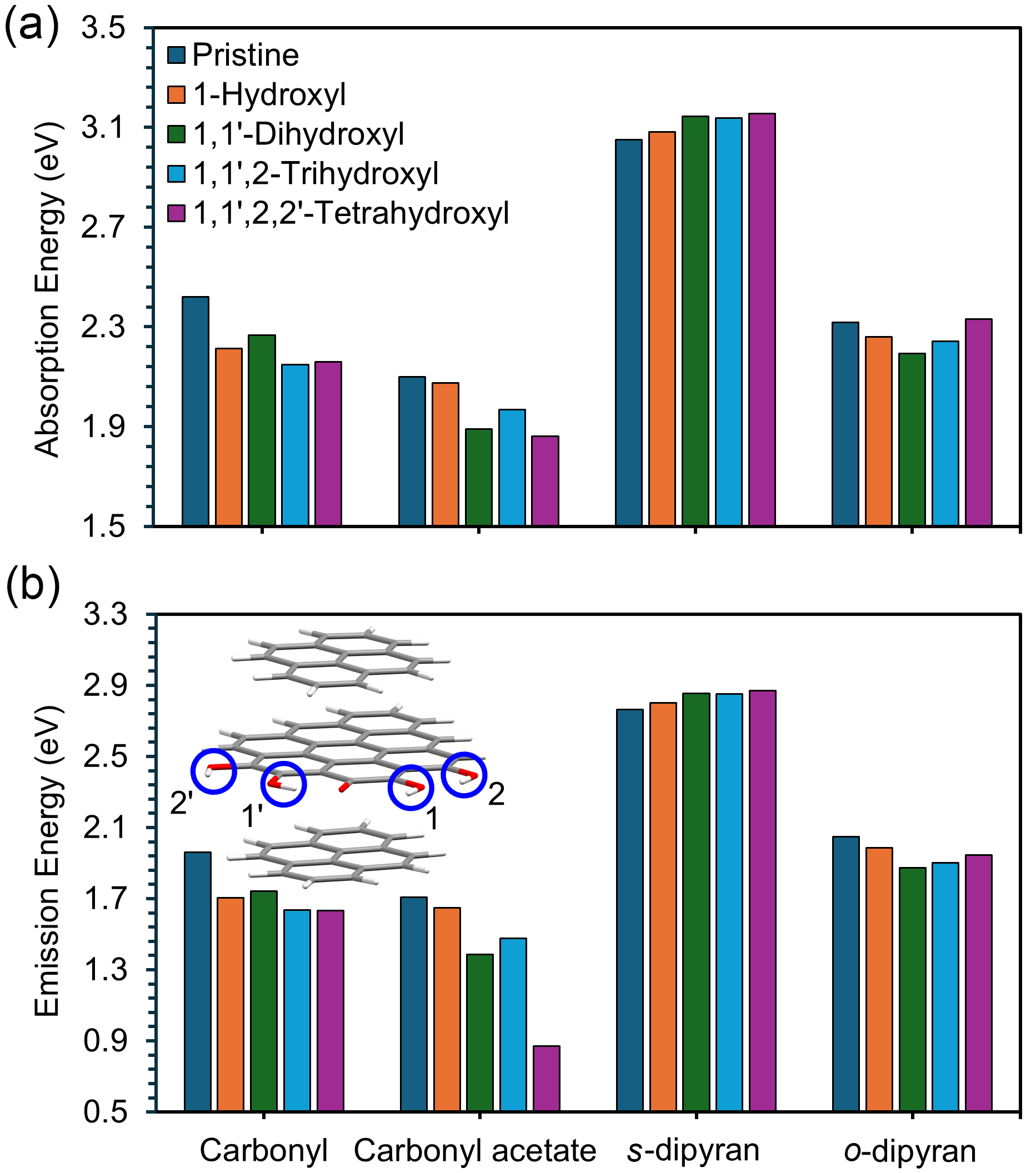}
 \caption{Absorption (a) and emission (b) energies (eV) of pristine and functionalized carbon dot models with varying numbers of hydroxyl groups. The presence of dominant surface defects reduces the influence of additional hydroxyl groups on the vertical transition energies.}
 \label{fgr:Figure2}
\end{figure*}
However, given the chemical diversity of surface defects on CDs, it is important to investigate whether spectator defects can still modulate these optical responses when present alongside such dominant defects.  In fact, the interaction between pairs of defects in model CDs has been found to provide desirable behavior in some cases.\cite{Sarkar2016,Umami2022,Jabed2021,Santika2022}
To investigate this, we analyzed the impact of spectator groups on the excitation and emission energies in the presence of dominant defects. Specifically, hydrogen atoms adjacent to carbonyl, carbonyl acetate, \textit{s}-dipyran, and \textit{o}-dipyran deefcts were sequentially replaced with hydroxyl groups, varying from one to four substitutions (see inset of Figure \ref{fgr:Figure2}). Figure \ref{fgr:Figure2} shows that, in most cases, increasing the number of –OH groups has a minimal effect on excitation energies when a dominant defect is present. For the carbonyl group, excitation energies exhibit a moderate redshift of 0.22 eV, with a similar trend in emission energies. A comparable trend is initially observed for carbonyl acetate. However, a notable deviation occurs when four adjacent hydrogens are replaced by hydroxyl groups in this system, the emission energy drops significantly by 0.89 eV. This larger shift arises from excited state intramolecular proton transfer (ESIPT) between one of the hydroxyl groups and a neighboring carbonyl.  These ESIPT reactions are well known in organic photochemistry,\cite{Sedgwick2018,Warburton2022} and it is not surprising that they would arise in CDs, as well.

Overall, these results suggest that while dominant defects predominantly govern the PL behavior of CDs, in specific cases, such as with carbonyl acetate, spectator groups may exert a more substantial effect when the interacting defects introduce new pathways for structural reorganization in the excited state. Nevertheless, in most cases, the influence of the spectator groups is negligible, and the PL properties are largely dictated by the dominant surface groups. This finding supports the notion that specific surface defects can introduce trap states, resulting in experimentally-observed excitation-independent fluorescence.\cite{Ding2016} However, it is not obvious that such states would be strongly localized to a single defect site.  Instead, a defect may effectively lower the energy of a particular sp$^2$ subdomain, creating a trap that is not strongly localized.
\par
\begin{figure*}[!htbp]
 \centering
 \includegraphics[height=16.5 cm]{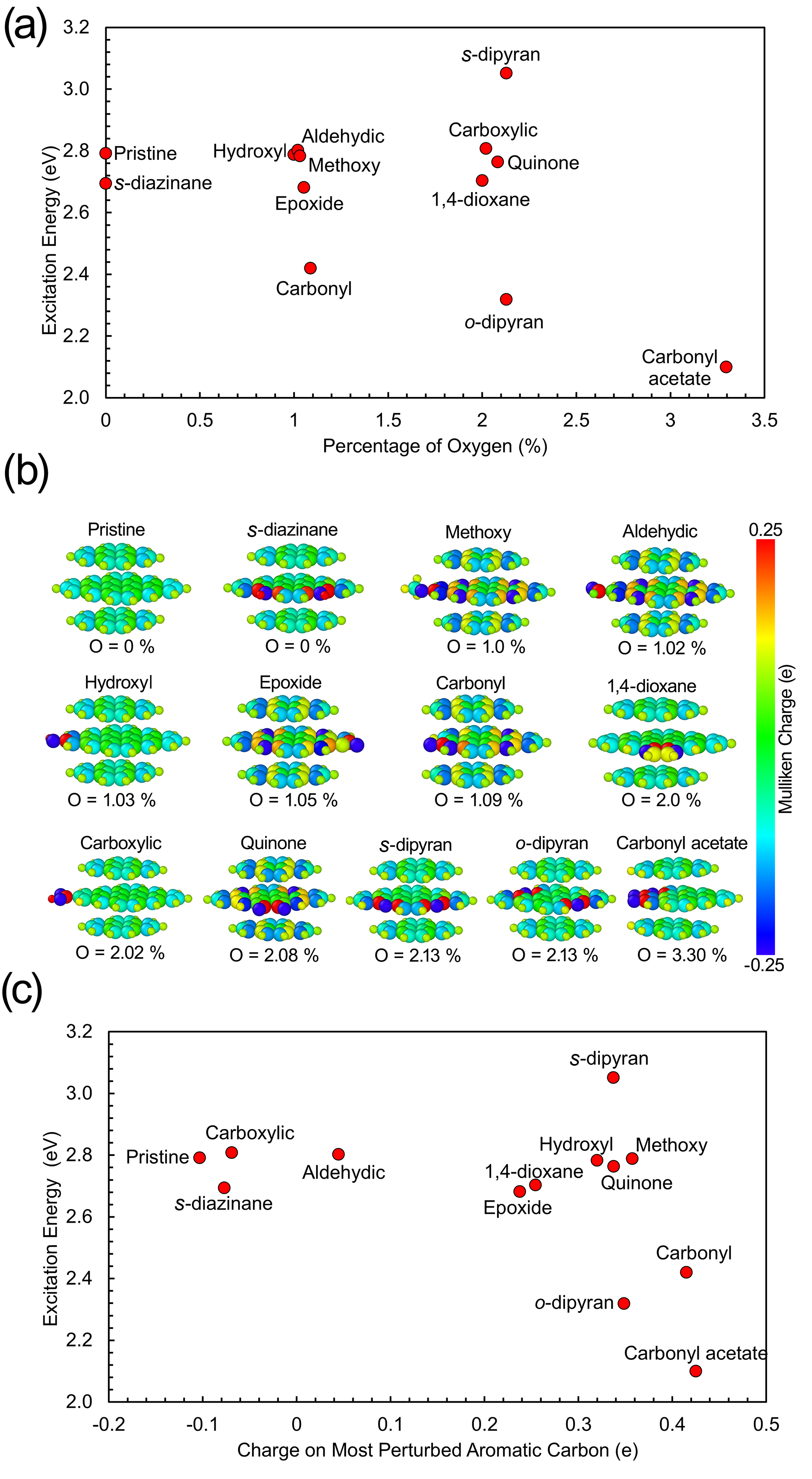}
 \caption{(a) Relationship between excitation energy and oxygen content (\%) for CD models with different surface defects. Increasing oxygen content generally lowers the excitation energy, corresponding to a smaller optical gap and a red-shift in absorption. (b) Ground-state Mulliken partial atomic charge distributions for representative CD models containing various surface defects. Atoms are color-coded by charge: green denotes near-neutral atoms, red indicates positive charge accumulation, and blue indicates negative charge accumulation. (c) Correlation between excitation energy (eV) and the ground-state charge (e) on the most perturbed aromatic carbon atom for CDs functionalized with different surface defects. Each point represents a model with a single defect type, illustrating how local charge perturbations modulate excitation energy.}
 \label{fgr:Figure3}
\end{figure*}
To better understand why certain type of defects dominate the optical response of CDs, it is important to look beyond their local bonding motifs and consider broader structural features that modulate their electronic behavior. Experimental studies have suggested that the oxygen content of surface groups plays a critical role in determining the position and intensity of absorption features in CDs, with higher oxygen incorporation often associated with smaller optical gap and red-shifted transitions.\cite{Zheng2011,Mei2010,Xu2013,Li2016, Srivastava2018, Liu2019b, Ding2016} In addition, several theoretical works suggest that oxygen-containing defects shift spectra of CDs to the red.\cite{Hola2014b, Sudolská2015, Feng2017, Ershov2025, Zhao2014} Motivated by these observations, we ask whether the degree of oxidation, quantified as the percentage of oxygen atoms within each surface group, can serve as a predictive metric for excitation energies. Specifically, we investigate whether increased oxygen content, either directly or through its impact on charge redistribution, consistently correlates with a narrowing of the optical gap in functionalized CD models. As shown in Figure \ref{fgr:Figure3}(a), this trend is clearly reflected in our simulations. Surface groups with higher oxygen content, such as carbonyl acetate ($\sim$3.3\%), exhibit lower excitation energies, consistent with red-shifted absorption and a more delocalized electronic structure. In contrast, less oxidized groups like hydroxyl display higher excitation energies near 2.7 eV, corresponding to a larger optical gap and blue-shifted absorption. Notably, deviations from this trend, such as the higher excitation energy observed for \textit{s}-dipyran despite its relatively high oxygen content, underscores the complex interplay between chemical structure, oxidation type, interactions with the $\pi$-system of the CD core, and electronic properties in determining the optical gap and emission characteristics. Overall, these results suggest that oxygen content serves as a useful, though not exclusive, indicator of a group's impact on the optoelectronic properties of CDs.
\par
Building on the observed correlation between oxygen content and excitation energy, we next investigate whether these trends can be rationalized in terms of changes in the underlying charge distribution. Figure \ref{fgr:Figure3}(b) visualizes the ground state Mulliken partial atomic charges for a series of functionalized CD models, with atoms color-coded to reflect their local charge states. More oxidized atoms are displaying higher positive charges (red and yellow regions) and less oxidized atoms are showing negative or neutral charges (blue and green regions). As oxygen content increases, a clear shift in the charge distribution is observed. For instance, in highly oxidized defects like carbonyl acetate, the carbon atoms near oxygen exhibit significant positive charges, suggesting strong polarization effects. This charge redistribution likely contributes to the reduction in the optical gap, as reflected in the lower excitation energies seen in these highly oxidized defects. On the other hand, in surface defects with lower oxygen content, such as hydroxyl and \textit{o}-diazinane, there is less accumulation of positive charge on the carbon atoms, which corresponds to a larger optical gap and higher excitation energies, leading to bluer emissions. The case of the carboxylic acid defect is particularly notable. Although it has a relatively high oxygen content, the strongly oxidized carbon is not part of the aromatic system of the CD core. This localized oxidation may explain why carboxylic acid defects maintain relatively higher excitation energies compared to other highly oxidized groups, as the integrity of the aromatic system remains largely intact, preserving the electronic structure responsible for a larger optical gap. To complement the atomic-centered partial charge analysis, electrostatic potential (ESP) isosurfaces were also generated, and results are provided in Figure S4. The ESP maps reveal that regions of negative ESP are located around defect sites, which is consistent with the charge distributions inferred from the partial charges. 
\par
To further explore the connection between partial atomic charges and excitation energies, Figure \ref{fgr:Figure3}(c) plots the excitation energy against the charge on the most perturbed aromatic carbon for each defect type on CDs. This figure provides a more granular view of how specific changes in charge distribution, especially on the aromatic carbons, correlate with the observed electronic properties. The plot shows that, as the charge on the most perturbed aromatic carbon becomes more positive, there is a general trend of decreasing excitation energy, which aligns with the earlier observation that increased oxygen content leads to red-shifted emissions. For example, carbonyl acetate, with a charge of approximately 0.43e on the most perturbed carbon, exhibits the lowest excitation energy among the samples. This is consistent with its high oxygen content and strong charge polarization effects, as seen in Figure \ref{fgr:Figure3}(a). Conversely, surface groups like \textit{s}-diazinane, which show minimal perturbation in charge on the aromatic carbon, maintain higher excitation energies. This further supports the notion that less oxidized defects preserve the electronic structure of the CDs, resulting in a larger optical gap and bluer emissions, and the extent of charge perturbation, particularly on aromatic carbons, is a key factor influencing the excitation energies in CDs. 
\par
Although the overall trend supports the role of oxygen-induced polarization in modulating excitation energies, there are notable exceptions that caution against over-reliance on partial charge analysis. For example, the difference in charge distribution between \textit{s}-diazinane and \textit{o}-dipyran appears relatively small in Figure \ref{fgr:Figure3}(b), yet these systems exhibit significantly different excitation energies. This discrepancy suggests that changes in electron density alone may not fully account for the differences in optical behavior. Instead, additional factors such as conjugation pathways, orbital localization, protonation state, and dot size etc., likely play important roles in determining not only the excitation energy but also the nature of the excitation itself. To address this, we next examine how these structural and electronic variables influence the type of excitations observed in CDs, including transitions of localized, charge-transfer, and $n$→$\pi^*$ character.

\subsection{Determinants of Excitation Character}
\begin{figure*}[bt]
 \centering
 \includegraphics[height=8.4cm]{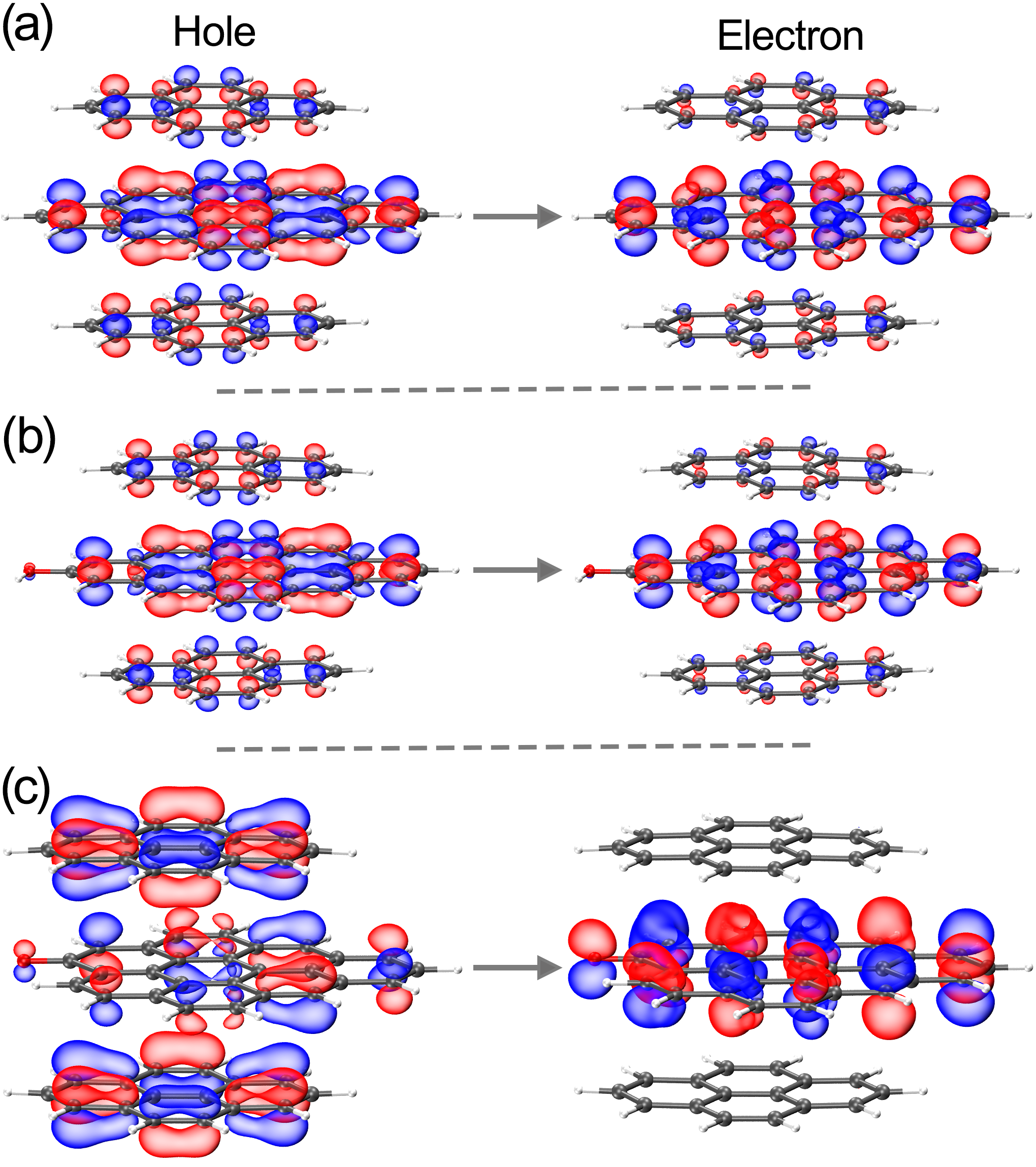}
 \caption{Natural transition orbitals (NTOs) for the lowest vertical singlet S$\textsubscript{0}$\textrightarrow S$\textsubscript{1}$ excitation of (a) a pristine model and a model functionalized with (b) hydroxyl and (c) carbonyl functional groups. The plots use an isovalue of 0.03 au.}
 \label{fgr:Figure4}
\end{figure*}

While the previous analysis focused on how oxygen content and local charge redistribution influence excitation energies, these descriptors alone do not capture the qualitative nature of the excitations. A more complete understanding requires examining the nature of the excitations themselves. To address this, we analyzed the natural transition orbitals (NTOs) which are modified canonical orbitals that aim to simplify the representation of an excited state transition by focusing on one or two key orbital pairs, thereby highlighting the most significant aspects of the transition.\cite{Martin2003} Figure \ref{fgr:Figure4} presents the NTO plots for selected model CDs of Figure \ref{fgr:Figure1}, while Figure S5 shows the plots for the remaining surface groups. A comparison of NTOs between functionalized CDs and the pristine model reveals that for the first electronic transition (S$\textsubscript{0}$\textrightarrow S$\textsubscript{1}$), the spectator defects, which do not dramatically alter the excitation and emission energies, maintain the $\pi$ to $\pi$* character of the transition observed in the pristine model. This transition is locally excited and primarily involves a redistribution of electron density within the central layer. However, the character of this transition is sensitive to the size of the CD model used in the study, which will be discussed in detail below. In this group, the electron-hole pairs are not significantly separated, and the degree of electron transfer from the holes is weak because of the strong interaction between electrons and holes in electronic transitions. Similar trends have been observed in single-layer perylene-based CD models.\cite{Kundelev2020} Depending on their position, some of these defects, such as carboxylic and epoxide, can serve as nonradiative centers for electron-hole recombination.\cite{Loh2010, Feng2017} 

Conversely, the dominant surface defects, which exhibit significant redshifts in excitation and emission energies, display mixed behavior (Figure \ref{fgr:Figure4}). The \textit{o}-dipyran system preserves the local $\pi$ to $\pi$* character involving the central layer, while carbonyl and carbonyl acetate demonstrate a charge transfer $\pi$\textrightarrow$\pi$* characteristic, where the charge transfer occurs from the outer layer to the central layer. This highlights the substantial impact that a single defect type can have on the photophysical behavior of CDs. Zboril \textit{et al.}\cite{Hola2014b} demonstrated that the functional groups such as carbonyl preferentially localize electrostatic charges towards the periphery of the layer, thereby acting as emissive surface traps. Such findings underscore the importance of the ability to fine-tune these properties through surface modification, particularly by adjusting the degree of oxidation, which offers a strategic approach to tailoring the photophysical characteristics of CDs.
\par
To further elucidate the impact of surface chemistry on photophysical behavior, we examined the oscillator strengths associated with the lowest-energy absorption and emission transitions (Figure S6). Strong $\pi$\textrightarrow$\pi$* transitions, such as those in pristine, hydroxylated, and methoxy functionalized systems, exhibit large oscillator strengths (\textit{f} $\approx$ 0.4–0.5 a.u), indicating intense radiative transitions. In contrast, systems containing strongly oxidizing groups, including carbonyl, carbonyl acetate, and quinone, show markedly reduced intensities (\textit{f} < 0.1 a.u), consistent with their charge-transfer character and lower orbital overlap. These trends indicate that chemical oxidation not only red-shifts the spectra but also suppresses radiative strength by reducing transition dipole coupling between frontier orbitals.
\par
Carbonyl-containing defects, such as carboxylic acid, on the surface of CDs can undergo deprotonation, significantly affecting their excitation and emission energies. This property of CDs has been investigated experimentally for the development of pH sensors, with various mechanisms proposed to explain the impact of pH variations.\cite{Chen2020, Jiao2018, Hoang2024, John2024} These mechanisms include the extension of conjugation upon deprotonation, reversible transformations between azo and quinone structures, and variations in the basicity of nitrogen-containing functional groups. However, previous theoretical studies did not identify any substantial qualitative changes in the overall absorption spectra following the deprotonation of surface groups.\cite{Sudolská2015} We, therefore, examined the influence of protonation states on the excitation and emission energies of CDs containing carboxylic and hydroxyl (as phenol) groups. As presented in Table 1, an increase in pH, leading to the formation of carboxylate and phenolate anions, results in a red shift in both absorption and emission peaks compared to their protonated forms. The transition to phenolate causes a significant red shift in the excitation energy by 0.78 eV, while the emission energy decreases with a shift of 0.66 eV. Similarly, for the transition from carboxylic acid to carboxylate, the excitation energy decreases with a shift of 0.25 eV, and the emission energy exhibits a dramatic reduction from 2.39 eV to 0.77 eV (a shift of 1.62 eV). These observations align with experimental findings. Gruebele and co-workers reported that the fluorescence intensity of oxygen-containing CDs with both red and blue emissions gradually decreases as the pH increases from 2 to 12.\cite{Nguyen2020} More recent studies provide consistent evidence that deprotonation stabilizes lower-energy emissive states.\cite{Liu2023, Szapoczka2024, Zhou2024} Liu et al. observed that CDs exhibit red-shifted and more intense emission in basic media, attributed to deprotonation and formation of surface charge traps.\cite{Liu2023} Szapoczka et al. reported a ratiometric response, with the longer-wavelength band gaining intensity as pH increases, reflecting growth of deprotonated surface states. \cite{Szapoczka2024} Time-resolved measurements further confirmed that deprotonation stabilizes n\textrightarrow$\pi$* type transitions, yielding red-shifted emission and longer lifetimes.\cite{Zhou2024} These experimental results collectively support the results given in Table 1. Additionally, pH variations influence the types of electronic transitions; for example, as shown in Figure S7, the deprotonation of a carboxylic group in CDs shifts the transition from a $\pi$\textrightarrow$\pi$* to an n\textrightarrow$\pi$* transition in the corresponding carboxylate form. We also examined the solvent dependence of the protonated and deprotonated models listed in Table 1 using the same series of solvent environments described above. The calculated absorption and emission energies (Figure S8) show that the protonated defects (carboxylic and hydroxyl) exhibit only minor solvent-induced changes (< 0.03 eV). In contrast, the deprotonated defects exhibit pronounced solvent-dependent behavior. For the carboxylate, the emission energy increases markedly from 0.77 eV in the gas phase to approximately 2.54 eV in polar solvents, accompanied by a modest absorption shift of about 0.21 eV. Similarly, for the phenolate, both absorption and emission energies show notable blue shifts of roughly 0.45 eV and 0.52 eV, respectively. These substantial shifts reflect strong dielectric stabilization of the anionic ground states in solution. We emphasize that these solvent-dependent shifts for fixed charge states are distinct from the pH-dependent trends discussed earlier (i.e., protonated vs. deprotonated species within the same medium).

\begin{table}
  \caption{Comparison of excitation and emission energies of carboxylic and hydroxyl-containing CD models in protonated and deprotonated states.}
  \label{tbl:pH}
  \centering
  \begin{tabular}{lcc}
    \hline
    Group        & Excitation (eV) & Emission (eV) \\
    \hline
    Carboxylic   & 2.81           & 2.39         \\
    Carboxylate  & 2.55           & 0.77         \\
    Hydroxyl     & 2.78           & 2.38         \\
    Phenolate    & 2.00           & 1.72         \\
    \hline
  \end{tabular}
\end{table}
\par
\begin{figure*}[ht]
 \centering
 \includegraphics[height=12.0 cm]{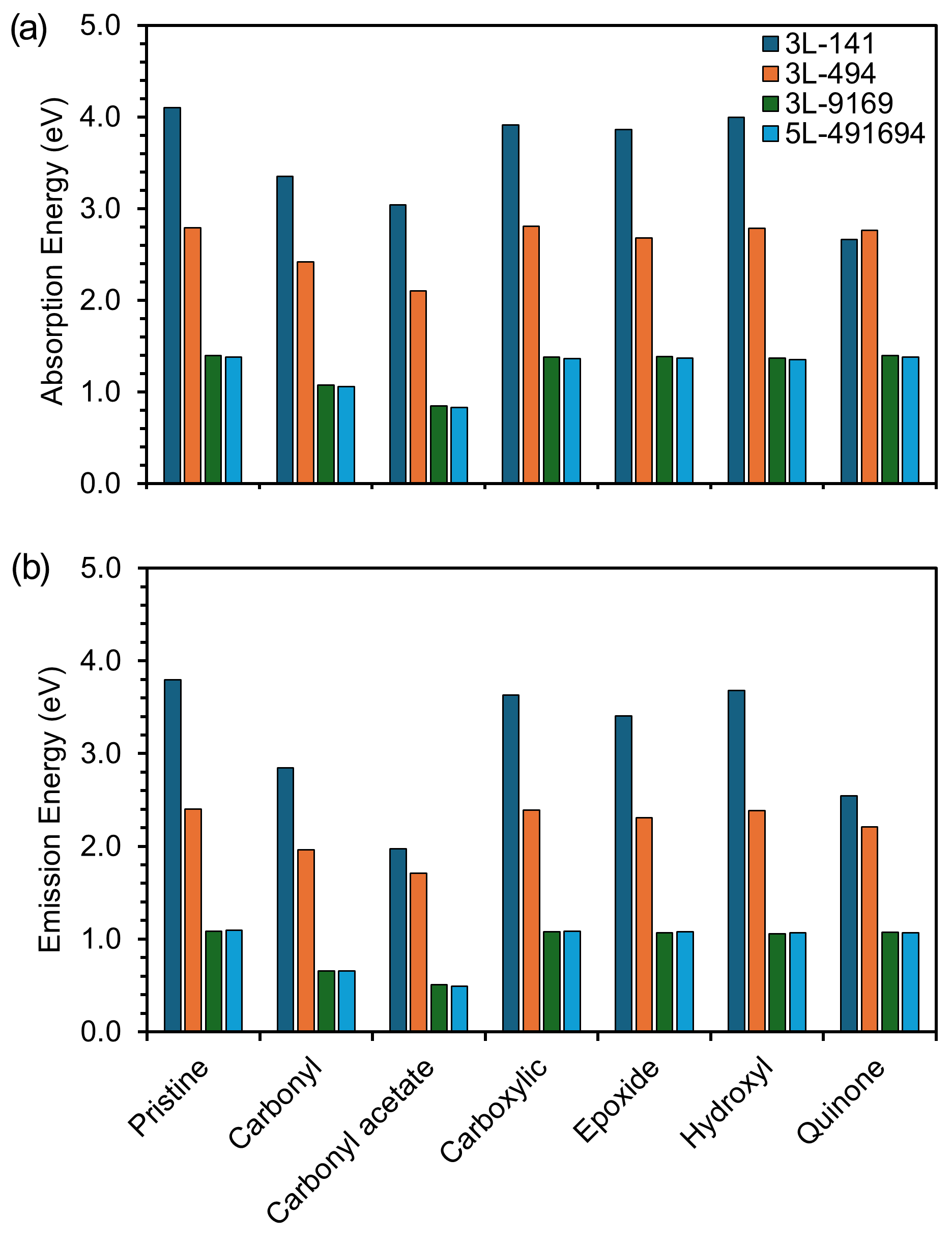}
 \caption{Absorption (a) and emission (b) energies (eV) of variously sized CD models containing oxygen-bearing defects reveal that increasing the size leads to a redshift in the spectrum. However, the size of the functionalized layer plays a crucial role, while additional layers contribute negligibly.}
 \label{fgr:Figure5}
\end{figure*}

We conclude this  section by examining the impact of core size and number of layers on the absorption and emission energies of CDs. We employed three CD models consisting of three layers (3L) and one model with five layers (5L) of sp$^2$-hybridized benzene rings. The models are labeled as 3L-$mnm$ and 5L-$lmnml$, where $l$, $m$, and $n$ represent the number of aromatic cyclic rings in each layer. These systems were functionalized with selected oxygen-containing defects: carbonyl, carbonyl acetate, carboxylic, epoxide, hydroxyl, and quinone. The calculated absorption and emission energies for these models are presented in Figure \ref{fgr:Figure5}. A comparative analysis of the energies reveals that both excitation and emission wavelengths exhibit a redshift as the size of the sp$^2$ carbon domain increases, consistent with previous experimental findings.\cite{Wang2021,Sk2014} However, when comparing the larger 3L-9169 model to the 5L model, which has an inner layer of a similar sp$^2$ size but two additional outer layers, there is negligible to no change in the transition energies. This observation aligns with our previous computational findings,\cite{Bian2023} indicating that further increasing the number of outer layers, without altering the central sp$^2$ domain, does not significantly affect the transition energies.
\begin{figure*}[ht]
 \centering
 \includegraphics[height=12.0 cm]{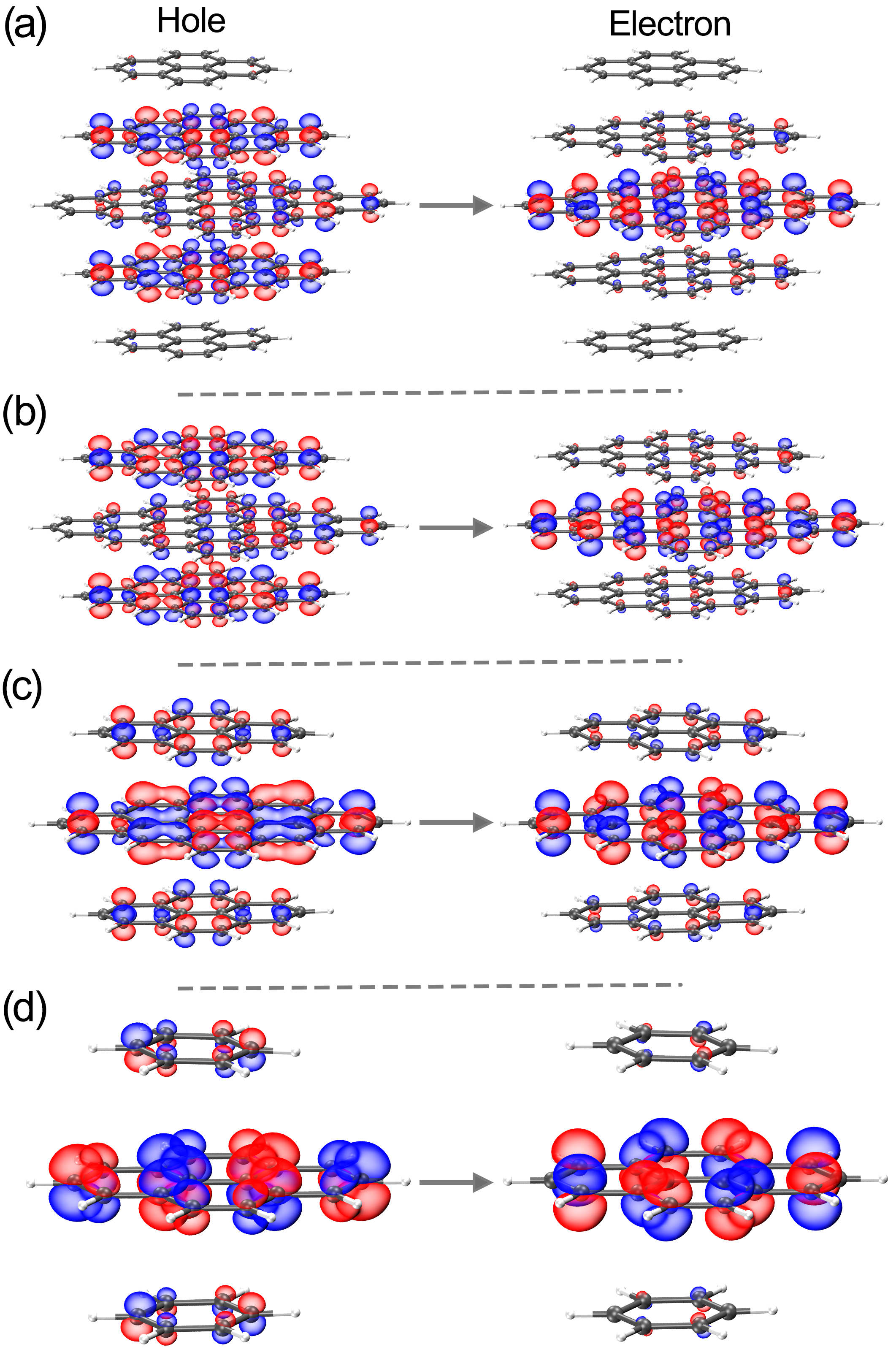}
 \caption{Natural transition orbitals (NTOs) for the lowest vertical singlet S$\textsubscript{0}$\textrightarrow S$\textsubscript{1}$ excitation of pristine CD models using (a) 5L-491694, (b) 3L-9169, (c) 3L-494 and (d) 3L-141 models. The plots use an isovalue of 0.03 au.}
 \label{fgr:Figure6}
\end{figure*}
\par
The relative comparison of NTOs between the 3L-494 and 3L-9169 models suggests that increasing the sp$^2$ domain also alters the excitation character of the transition. For example, in the pristine 3L-494 system, as shown in Figure \ref{fgr:Figure6}c, the transition primarily involves the central layer. In contrast, the 3L-9169 model (Figure \ref{fgr:Figure6}b) demonstrates a charge-transfer character, with the hole delocalized across the outer layers and the electron localized in the central layer. A similar change in excitation character is observed in other functionalized systems, where transitions in the 3L-494 model  predominantly involve the central layer. Notably, the carbonyl-decorated 3L-141 (Figure \ref{fgr:Figure6}d) model exhibits excitations localized exclusively within the central layer, a behavior distinct from that observed in all other models considered. The NTO analysis of the 5L model (Figure \ref{fgr:Figure6}a) further reveals that the outer layers do not contribute to the hole and electron distributions, exhibiting a $\pi$\textrightarrow$\pi^*$ charge-transfer transition character similar to the 3L-9169 model, involving only the central and adjacent layers. These findings underscore the significance of the sp$^2$ core size in determining both the transition energies and the nature of the electronic transitions for CDs, even in the absence of any defects or dopants. These trends are in line with our earlier work on related models \cite{Bian2023}, where large conjugated domains showed the strongest size dependence, interrupted-conjugation motifs showed weaker size dependence, and layering effects were comparatively small. Together, these results show that while the absolute excitation energies depend on the core size, the underlying relationships between surface chemistry, and optical response remain consistent once the conjugated domain reaches moderate dimensions.

\subsection{Conformational Effects on Excitation Behavior}
Recent experimental studies have reported transient fluctuations in the polarization and emission intensities of CDs in single-dot emission microscopy.\cite{Gomez2025} These fluctuations may be attributable to conformational dynamics involving the chromophoric regions, either through global rotations of the dot or rotations and/or displacements of individual chromophore-bearing layers.\cite{Gomez2025} Such dynamics may directly impact the photoluminescence of a single chromophore or introduce/remove opportunities for energy transfer into non-emissive (dark) states.\cite{Gomez2025} To investigate the effects of such intra-dot dynamics, we studied model three-layer CDs where the outer layers are functionalized with either identical or distinct surface groups. Our objective was to determine how the twisting and sliding of the layers in relation to one another alters the nature of the electronic excitations and the corresponding excitation energies.  Though we focus here on excitation energies for computational simplicity, it is reasonable to extrapolate that similar changes in emission energies would also be observed.
\par
\begin{figure*}[ht]
 \centering
 \includegraphics[height=4.95 cm]{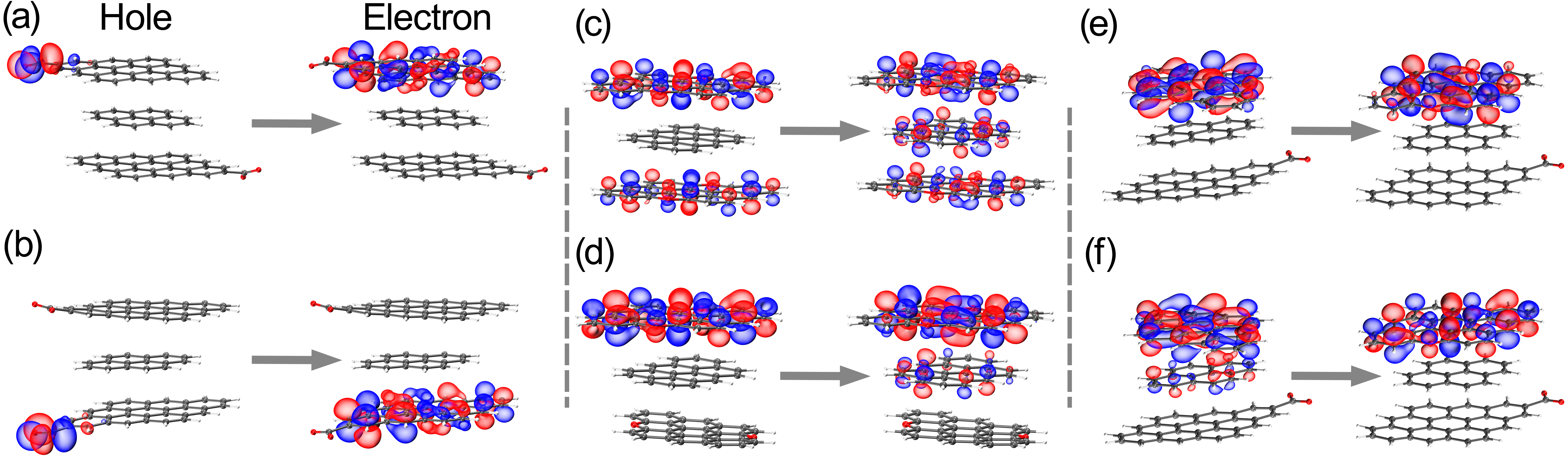}
 \caption{NTOs of the first single excited states of model CDs illustrating the effect of rotating the carboxylate- or \textit{p}-dipyran-functionalized outer layers. (a, b) CDs with carboxylate groups on both outer layers: (a) carboxylates positioned on opposite sides; (b) bottom layer rotated so both carboxylates are on the same side. (c, d) CDs with \textit{p}-dipyran groups on both outer layers: (c) pyran groups aligned on the same side; (d) bottom layer twisted by 90° to place pyran groups on opposite sides. (e, f) CDs with asymmetric functionalization: pyran on top and carboxylate on bottom layers; (e) both groups aligned on the same side; (f) pyran layer twisted by 90°. Twisting alters the spatial character and localization of the excitations, with symmetric functionalization showing sensitivity in excitation type, while asymmetric systems remain dominated by the pyran group. These conformational effects provide mechanisms for experimentally observed polarization fluctuations and emission intermittency in CDs. The plots use an isovalue of 0.03 au.}
 \label{fgr:Figure7}
\end{figure*}

Figure \ref{fgr:Figure7}a presents the NTOs for a model CD with carboxylate groups on both outer layers. In this configuration, the carboxylates are positioned on opposite sides of the dot. The excitation is of n\textrightarrow$\pi^*$ character and is predominantly localized on the outer layers. When the bottom layer is rotated such that both carboxylate groups point in the same direction (Figure \ref{fgr:Figure7}b), the character of the excitation remains largely unchanged. However, as shown in Table \ref{cdot_conf}, the corresponding excitation energy shifts by 0.27 eV, with the more stable initial conformation exhibiting a higher excitation energy.  This change in excitation energy is accompanied by a significant 69.5° reorientation of the dipole vector and a 31\% reduction in oscillator strength. These findings suggest that the relative orientation of functionalized surface layers can influence both the excitation energy and the electronic character of CDs.  The initial geometry is 0.38 eV more stable at the ground-state optimized geometry.
\par
Next, we examined the effect of twisting when both outer layers are functionalized with \textit{o}-dipyran groups. In the configuration where the pyran groups are aligned on the same side (Figure \ref{fgr:Figure7}c), the hole is delocalized over both outer layers, while the electron is delocalized across all three layers, indicating significant interlayer electronic coupling. Upon rotating one of the outer layers by 90° so that the pyran groups are on opposite sides (Figure \ref{fgr:Figure7}(d)), the character of the excitation changes dramatically. The excitation becomes more localized, confined primarily to the layer containing one of the pyran groups. The parallel conformation (Figure \ref{fgr:Figure7}c) is slightly more stable by 0.042 eV (Table \ref{cdot_conf}), and the change in excitation energy is minimal (0.022 eV).  However, the oscillator strength doubles in magnitude (99.7\% increase), and exhibits a slight 9.5° rotation. These results highlight that the relative orientation of functionalized surface layers can impact both the excitation character and the oscillator strength, even in the absence of a significant change in excitation energy.
\par
Next, we analyzed a hetero-functionalized CD, where one outer layer bears a carboxylate group and the other a pyran group. In the configuration where both groups point in the same direction (Figure \ref{fgr:Figure7}e), the excitation remains localized on the pyran-containing layer, with negligible involvement of the carboxylate side. A roughtly 90° twist of the pyran layer (Figure \ref{fgr:Figure7}f) does not significantly alter the excitation character, which remains localized on the pyran-functionalized layer. Upon twisting, a corresponding rotation of the transition dipole moment by 83.4\textdegree is observed.  Similar rotations in polarization angle were also observed in single-particle luminescence experiments.\cite{Gomez2025}  Small changes are also observed in the excitation energy:  a 0.03 eV downward shift is observed upon rotation.  Only a 0.08 eV difference in ground state energy is observed upon twisting. 

\begin{table}
  \caption{Relative ground-state energies ($\Delta E_{\text{S}_0}$), relative vertical excitation energies ($\Delta E_{\text{S}_0 \rightarrow \text{S}_1}$), changes in transition dipole orientation ($\Delta \theta_{\mu}$), and percentage changes in oscillator strength ($\Delta f$) for five different three-layer carbon dot architectures under configurational perturbations, pictured in Figures \ref{fgr:Figure7} and \ref{fgr:Figure8}. Each value is reported relative to the corresponding alternative configuration within the same system. The configurational changes include the rotation of a single layer (flake), sliding of a single flake between two domains of a second layer, and folding of about a flexible linker.}
  \label{cdot_conf}
  \centering
  \begin{tabular}{lccccc}
    \hline
    System & Configurational Change & $\Delta E_{\text{S}_0}$ & $\Delta E_{\text{S}_0 \rightarrow \text{S}_1}$ & $\Delta \theta_{\mu}$  & $\Delta f$ \\
          &               & (eV)             & (eV)                             & (°)                   & (\%)       \\
    \hline
    Carboxylate & Layer Rotation         & -0.38 &  0.27 &   69.5 & -31.2 \\
    \textit{p}-dipyran & Layer Rotation               &  0.04 &  0.022  & 9.5 &  99.7 \\
    Asymmetric & Layer Rotation          &  0.08 & -0.03  &  83.4 & -26.2 \\
    \ce{-CH=CH-} & Layer Slide           &  0.17 &  0.07  & 6.6 &  56.6 \\
    \ce{-CH2-CH2-CH2-} & Layer Folding &  0.85 &  0.12  & 27.2 &  50.4 \\
    \hline
  \end{tabular}
\end{table}

\par
These observation provide a potential molecular-level explanation for the experimentally observed fluctuations in polarization and emission intensities in CDs. The small energy differences between twisted and untwisted conformations suggest that such structural rearrangements are thermally accessible and likely to occur under ambient or excited-state conditions. In particular, twisting of the outer layers alters the spatial distribution and character of the electronic excitations, shifting from delocalized to more localized states depending on the nature and orientation of the surface groups. This modulation can influence exciton coupling between layers and potentially affect the rates of radiative versus non-radiative decay. Thus, the transient photophysical behavior observed experimentally, including blinking and polarization fluctuations, can be rationalized in terms of layer-dependent conformational motions that dynamically alter the electronic structure of the chromophore.
\par
We hypothesize that the experimentally observed fluctuations in CD emission may originate not only from torsional reorientations but also from sliding and folding motions of the surface-functionalized layers. Such dynamics can alter the electronic interactions between chromophoric and non-chromophoric layers, thereby modulating the charge transfer pathways and excitonic delocalization. Here we investigate whether these types of conformational changes can impact the energy and character of the S\textsubscript{1} state, offering additional plausible mechanisms for the observed dynamics. To this end, we studied model CDs with top-layer mobility induced through —CH=CH— and —CH\textsubscript{2}—CH\textsubscript{2}—CH\textsubscript{2}— linkers.
\par
\begin{figure*}[!ht]
 \centering
 \includegraphics[height=9.5 cm]{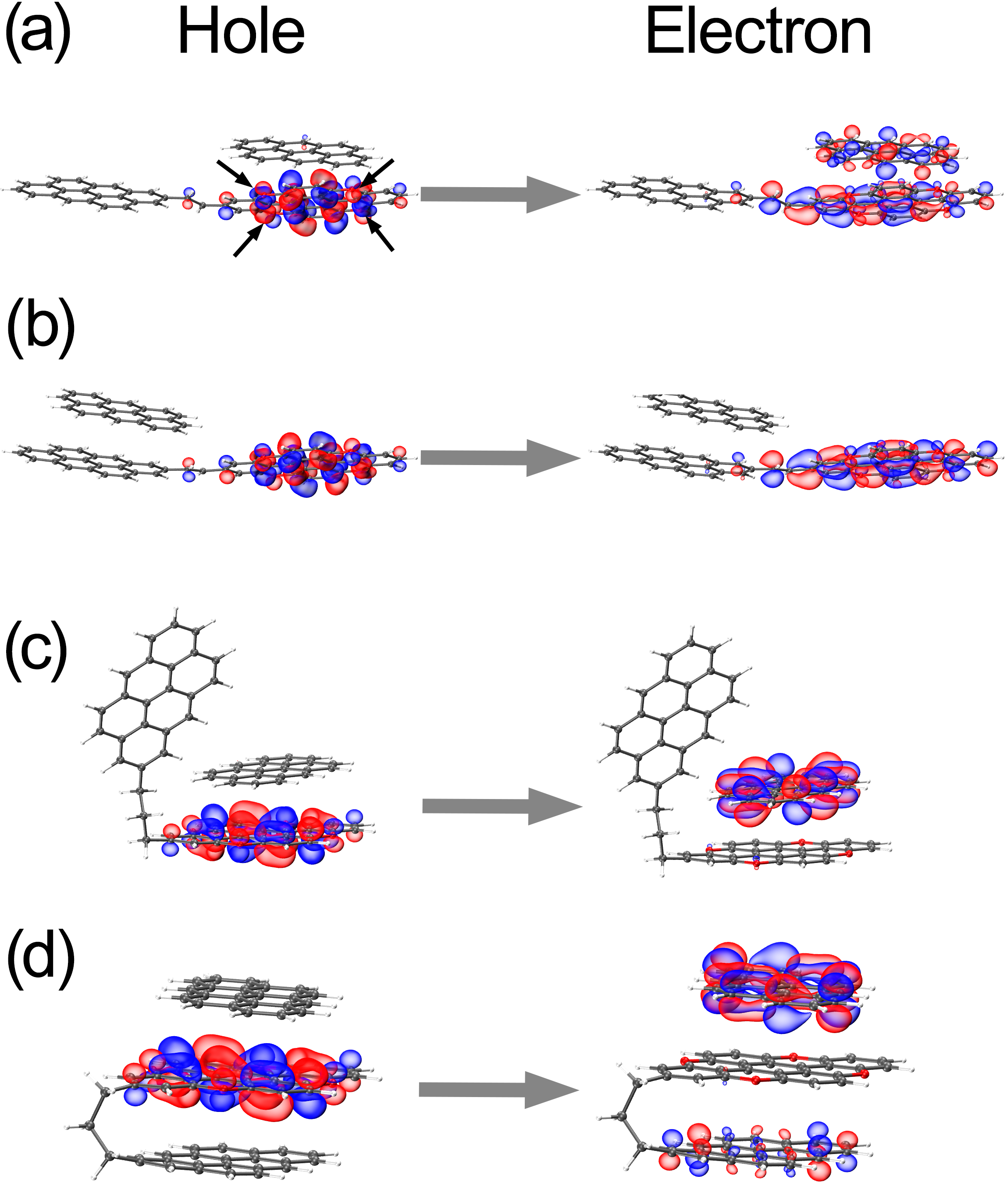}
 \caption{NTOs of the S\textsubscript{0} \textrightarrow S\textsubscript{1} transition for CD models with three flakes connected via flexible and rigid linkers, illustrating the effects of sliding and folding. (a, b) Rigid linker —CH=CH— with sliding of the third flake from the functionalized to the pristine side. The positions of the four functionalizing oxygen atoms are highlighted by black arrows. (c, d) Long linker —CH\textsubscript{2}—CH\textsubscript{2}—CH\textsubscript{2}— with (c) standing and (d) folded geometries. Excitation character and delocalization change significantly with layer motion, revealing how linker flexibility and layer positioning modulate interlayer charge transfer and excitation energy. The plots use an isovalue of 0.03 au.}
 \label{fgr:Figure8}
\end{figure*}

We first investigated a rigid linker system where two functionalized single layers (flakes)---a tetra-pyran-functionalized flake and a pristine flake---are connected via a —CH=CH— bridge. Figures \ref{fgr:Figure8}a and \ref{fgr:Figure8}b show the NTOs before and after sliding a third non-covalently attached flake from the functionalized to the pristine side. The ground-state energy difference between the two configurations is modest (0.17 eV), and the S\textsubscript{1} excitation energy changes by only 0.067 eV.  A ~50\% magnitude increase is predicted in the oscillator strength (Table \ref{cdot_conf}), without any significant change in the polarization angle. In both conformations, the S\textsubscript{0} \textrightarrow S\textsubscript{1} transition has $\pi$\textrightarrow$\pi^*$ character with the hole predominantly localized on the functionalized layer. However, in the conformation where the third layer sits above the functionalized layer (Figure \ref{fgr:Figure8}a), the electron is delocalized across both layers, indicating stronger interlayer electronic interactions. When the sliding layer moves to the pristine layer (Figure \ref{fgr:Figure8}b), the electron becomes localized. These findings demonstrate another facile structural change that can modulate both the oscillator strength and the interlayer coupling.
\par
Finally, we examined a folding motif using a more flexible —CH\textsubscript{2}—CH\textsubscript{2}—CH\textsubscript{2}— linker. Figures \ref{fgr:Figure8}c and \ref{fgr:Figure8}d show two configurations; a "standing" conformation (Figure \ref{fgr:Figure8}c), where the linked flake is orthogonal to the core, and a "folded" conformation (Figures \ref{fgr:Figure8}d), where the functionalized layer is sandwiched between two pristine flakes. The folded geometry is significantly more stable by 0.85 eV. The S\textsubscript{1} excitation energy also changes appreciably (0.12 eV), with both configurations exhibiting strong $\pi$\textrightarrow$\pi^*$ charge-transfer character. A 50\% increase in oscillator strength and a 27° change in polarization angle arise from this folding motion. Notably, the electron delocalization in the folded geometry extends across multiple layers. 
\par
These simplified models contain only a single chromophore.  It is clear that even in such small models, relatively facile conformational changes may trigger large changes in emission intensity and polarization direction.  Yet realistic CDs may be roughly 10 nm in diameter and may contain a large number of coupled chromophores.  One can imagine that the changes in electronic structure and transition dipole observed here could result in significant changes in inter-chromophore couplings and couplings between chromophores and dark trap states.  Such changes could lead to very complex dynamics that could also be responsible for experimentally-observed luminescence intensity and polarization fluctuations.

\section{Conclusions}
The findings presented in this study elucidate the significant influence of surface defects on the optical properties of CDs. As reported is past theoretical studies, strongly oxidizing defects, such as carbonyl and carbonyl acetate, are found to induce substantial red shifts in both absorption and emission energies by altering the electronic structure and excitation characters through enhanced charge distribution and orbital interactions. In contrast, less-oxidizing defects like hydroxyl and methoxy minimally perturb the optical properties, maintaining energy levels close to those of pristine CDs. Additionally, our results demonstrate that the photophysical characteristics of CDs can be fine-tuned by adjusting the degree of oxidation, with higher oxygen content generally leading to greater positive charge accumulation, narrower optical gaps, and red-shifted excitations. Yet the most-oxidizing defects are found to dominate the properties, while less-oxidizing spectator devects (such as carboxylate) play only a modest role.  Our results further highlight that excitation localization and charge-transfer behavior in CDs are highly sensitive to twisting, sliding, and folding motions between functionalized layers, providing mechanistic insight into their dynamic photophysical behavior. 
\par
The framework developed in this work establishes an actionable connection between surface chemistry and optical response in CDs. Although qualitative in nature, the computed trends in excitation energy, charge-transfer character, and protonation dependence closely align with recent experimental observations. This framework thus offers a physically grounded basis for interpreting spectroscopic data and guiding future combined experimental and computational investigations.

\begin{acknowledgement}
We gratefully acknowledge several helpful discussions with Martin Gruebele, Stephan Link, and Peter Rossky.  This work is supported by U.S. Department of Energy, Office of Science, Office of Basic Energy Sciences, under Award No. DE-SC0021643 and from the Institute for Advanced Computational Science, Stony Brook University. A part of the research was performed using computational resources sponsored by the Department of Energy's Office of Energy Efficiency and Renewable Energy and located at the National Renewable Energy Laboratory. The work also used Expanse GPU at the San Diego Supercomputer Center (SDSC) through allocation CHE140101 from the Advanced Cyberinfrastructure Coordination Ecosystem: Services and Support (ACCESS) program, which is supported by National Science Foundation grants \#2138259, \#2138286, \#2138307, \#2137603, and \#2138296.
\end{acknowledgement}

\begin{suppinfo}
Supporting figures and coordinates of optimized structures. 
\end{suppinfo}

\bibliography{CDots}

\end{document}